\documentclass[rnote]{aa}
\usepackage{graphicx}
\usepackage{txfonts}
\usepackage{natbib}
\usepackage{verbatim}
\usepackage{longtable}
\usepackage{lscape}
\usepackage{afterpage}
\usepackage{multirow}
\bibpunct{(}{)}{;}{a}{}{,}

\begin{document}

\title{Dehydrogenated polycyclic aromatic hydrocarbons and UV bump}

\author{G. Malloci\inst{1}
	\and
	G. Mulas\inst{1}
	\and
	C. Cecchi\textendash Pestellini\inst{1}
	\and
	C. Joblin\inst{2}
}

\institute{
Istituto Nazionale di Astrofisica~\textendash~Osservatorio Astronomico di 
Cagliari, Strada n.54, Loc. Poggio dei Pini, I\textendash09012 
Capoterra (CA) (Italy)
\textemdash\email{[gmalloci; gmulas; ccp]@ca.astro.it}
\and
Centre d'Etude Spatiale des Rayonnements, Universit\'e de Toulouse
\textendash CNRS, Observatoire Midi\textendash Pyr\'en\'ees, 9 Avenue du 
Colonel Roche, 31028 Toulouse cedex 04 (France)
\textemdash\email{christine.joblin@cesr.fr}
}
\date{Received 12 May 2008; accepted 05 August 2008}

\abstract
{Recent calculations have shown that the UV bump at about 217.5~nm 
in the extinction curve can be explained by a complex mixture of polycyclic 
aromatic hydrocarbons (PAHs) in several ionisation states.
Other studies proposed that the carriers are a restricted population made of 
neutral and singly\textendash ionised dehydrogenated coronene molecules 
(C$_{24}$H$_{n}$, $n\leq3$), in line with models of the hydrogenation state 
of interstellar PAHs predicting that medium\textendash sized species
are highly dehydrogenated.}
{To assess the observational consequences of the latter hypothesis
we have undertaken a systematic theoretical study of the electronic 
spectra of dehydrogenated PAHs. We use our first results to see
whether such spectra show strong general trends upon dehydrogenation.}
{We performed calculations using 
state\textendash of\textendash the\textendash art techniques 
in the framework of the density functional theory (DFT) to obtain the 
electronic ground\textendash state geometries, and of the 
time\textendash dependent DFT to evaluate the 
electronic excited\textendash state properties.}
{We computed the absorption cross\textendash section of the species 
C$_{24}$H$_{n}$ (n=12,10,8,6,4,2,0) in their neutral and cationic 
charge\textendash states. Similar calculations were performed for other PAHs
and their fully
dehydrogenated counterparts.}
{$\pi$\textendash electron energies are always found to be 
strongly affected by dehydrogenation. In all cases we examined, progressive 
dehydrogenation translates into a correspondingly progressive blue shift of 
the main electronic transitions. In particular, the $\pi\to\pi^*$ collective 
resonance becomes broader and bluer with dehydrogenation. Its 
calculated energy position is therefore predicted to fall in the gap 
between the UV bump and the far\textendash UV rise
of the extinction curve. Since this effect appears to be systematic,
it poses a tight observational limit on the column  
density of strongly dehydrogenated medium\textendash sized PAHs.}


\authorrunning{Malloci et al.}
\titlerunning{Dehydrogenated PAHs and UV bump}

\maketitle

\section{Introduction} \label{introduction}

First discovered by \citet{ste65a}, the absorption band centered at about
217.5~nm  ($\sim$5.7~eV) is the most intense interstellar extinction 
feature. With a relatively constant peak wavelength, the so\textendash called 
``UV bump'' has a Lorentz\textendash like profile \citep[e.g.,][]{sea79} whose 
width varies considerably from one line of sight to another \citep{fit07}.

Both the strength and the position of the 217.5~nm feature suggest that it 
originates in some form of carbonaceous material, known to display strong 
$\pi\to\pi^\star$ electron transitions in this range. A large number of carrier 
candidates have been proposed over the years, amongst which size\textendash restricted 
graphite particles \citep{ste65b}, fullerenes, C$_{60}$ in particular, 
\citep[e.g.,][]{bra91,igl04}, polycyclic aromatic hydrocarbons 
\citep[PAHs,][]{job92a,job92b}, coal\textendash like material \citep{pap96}, 
UV\textendash processed hydrogenated amorphous carbon particles \citep{men98}, 
nano\textendash sized hydrogenated carbon grains \citep{sch98}, and carbon 
onions \citep{chh03}.

Since each PAH displays strong $\pi\to\pi^\star$ transition in the 5.7~eV region 
\citep{job92a,job92b,dra03,mal04}, mixtures of PAHs must contribute to the 
217.5~nm bump. Recently, \citet{ccp08} showed that such mixtures can account 
for the entire bump, as well as its strength relative to the non\textendash linear far\textendash UV
rise, depending on the distribution of charge states in the PAH mixtures. The 
UV bump was also attributed to a  $\pi\to\pi^\star$ plasmon resonance from a very 
limited number of strongly dehydrogenated derivatives of coronene 
(C$_{24}$H$_{12}$), namely C$_{24}$H$_n$ and C$_{24}$H$_n$$^+$ with $n\leq3$. This 
assignment was made on the basis of DDA scattering calculations \citep{dul98}, 
supported by laboratory studies of the UV absorption in amorphous carbon 
\citep{dul04}. 
The same molecules were also proposed as carriers of some of the diffuse 
interstellar bands \citep{dul06b}. Indeed, models of the hydrogenation state 
of PAHs predict small and medium\textendash sized PAHs ($\sim$20-30 carbon atoms) to be 
highly dehydrogenated \citep{lep03}.

Since the effect of dehydrogenation on the electronic spectra of 
isolated PAHs is poorly characterized, we recently undertook a detailed
study of the electronic excitation properties of dehydrogenated PAHs,
using state\textendash of\textendash the\textendash art quantum\textendash chemical tools \citep{mal08}. As a part of
this more detailed work, we present in this paper our first results for the  
dehydrogenated derivatives of coronene (C$_{24}$H$_{12}$), and completely 
dehydrogenated perylene 
(C$_{20}$H$_{10}$) and ovalene (C$_{32}$H$_{14}$). We use them to
assess the specific case of the species introduced by \citet{dul06a}, 
whose computed spectral properties appear to be incompatible with the 
observed UV\textendash bump, and search for general trends in the 
electron properties of dehydrogenated PAHs. 

Technical details of the calculations, as well as the justification for
the choice of molecules considered, are given 
in Sect.~\ref{methods}. Results are presented and discussed in 
the astrophysical context in Sect.~\ref{results}, and our 
conclusions are reported in Sect.~\ref{conclusions}.

\begin{figure}[b!]

\begin{center}
\includegraphics[trim=3cm 3cm 3cm 3cm, clip, width=2.5cm]{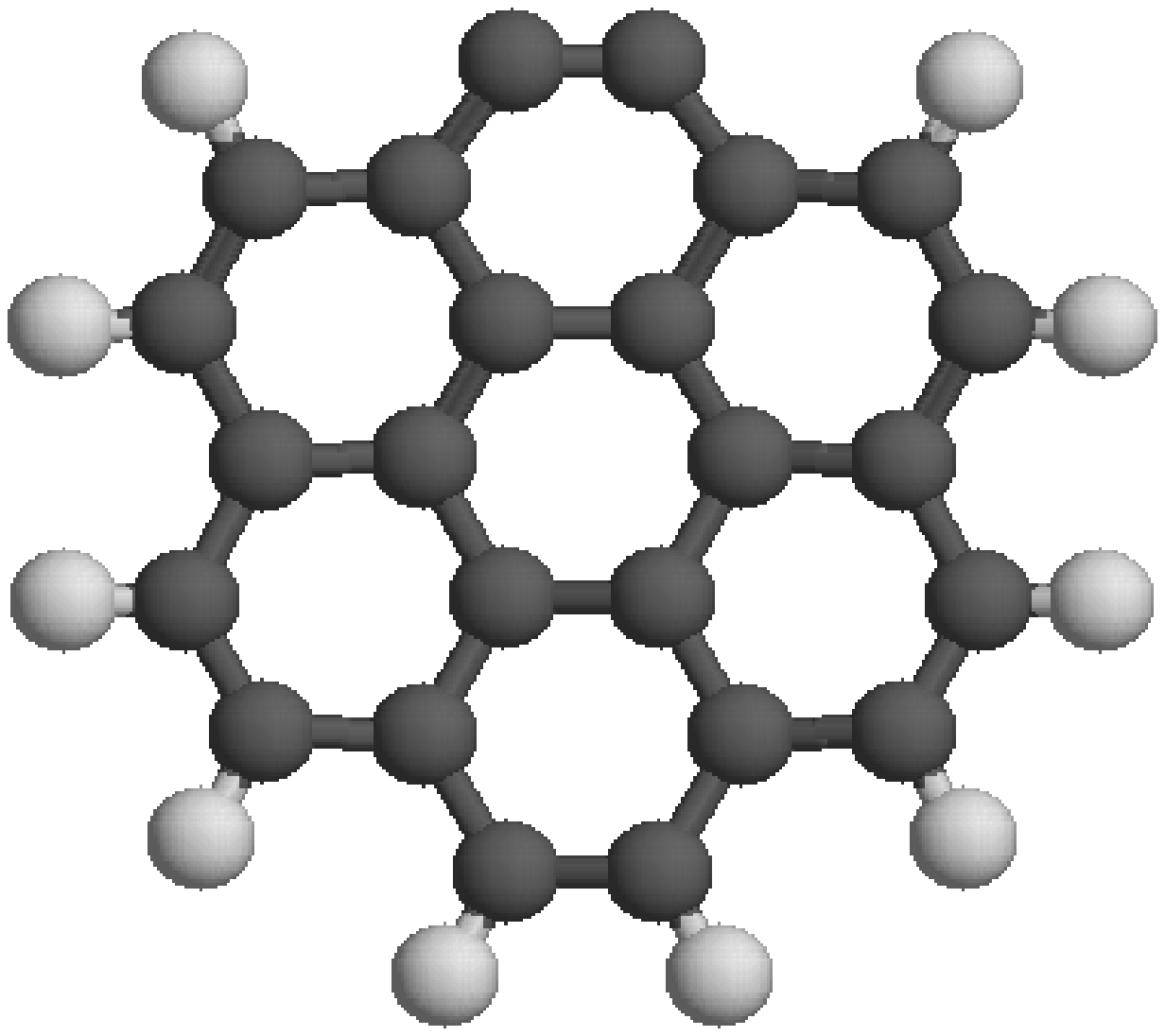}
\includegraphics[trim=3cm 3cm 3cm 3cm, clip,width=2.5cm]{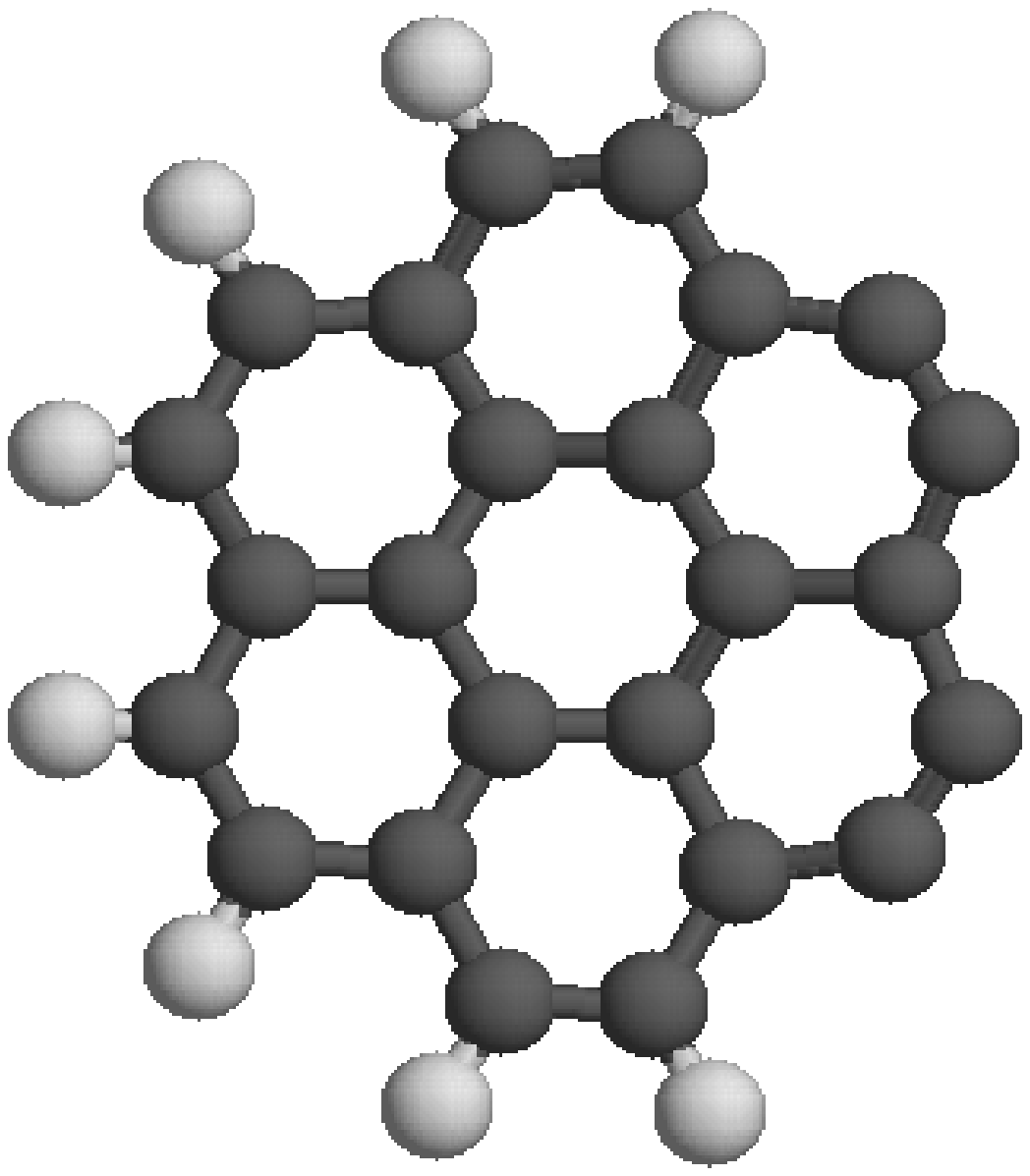}
\includegraphics[trim=3cm 3cm 3cm 3cm, clip,width=2.5cm]{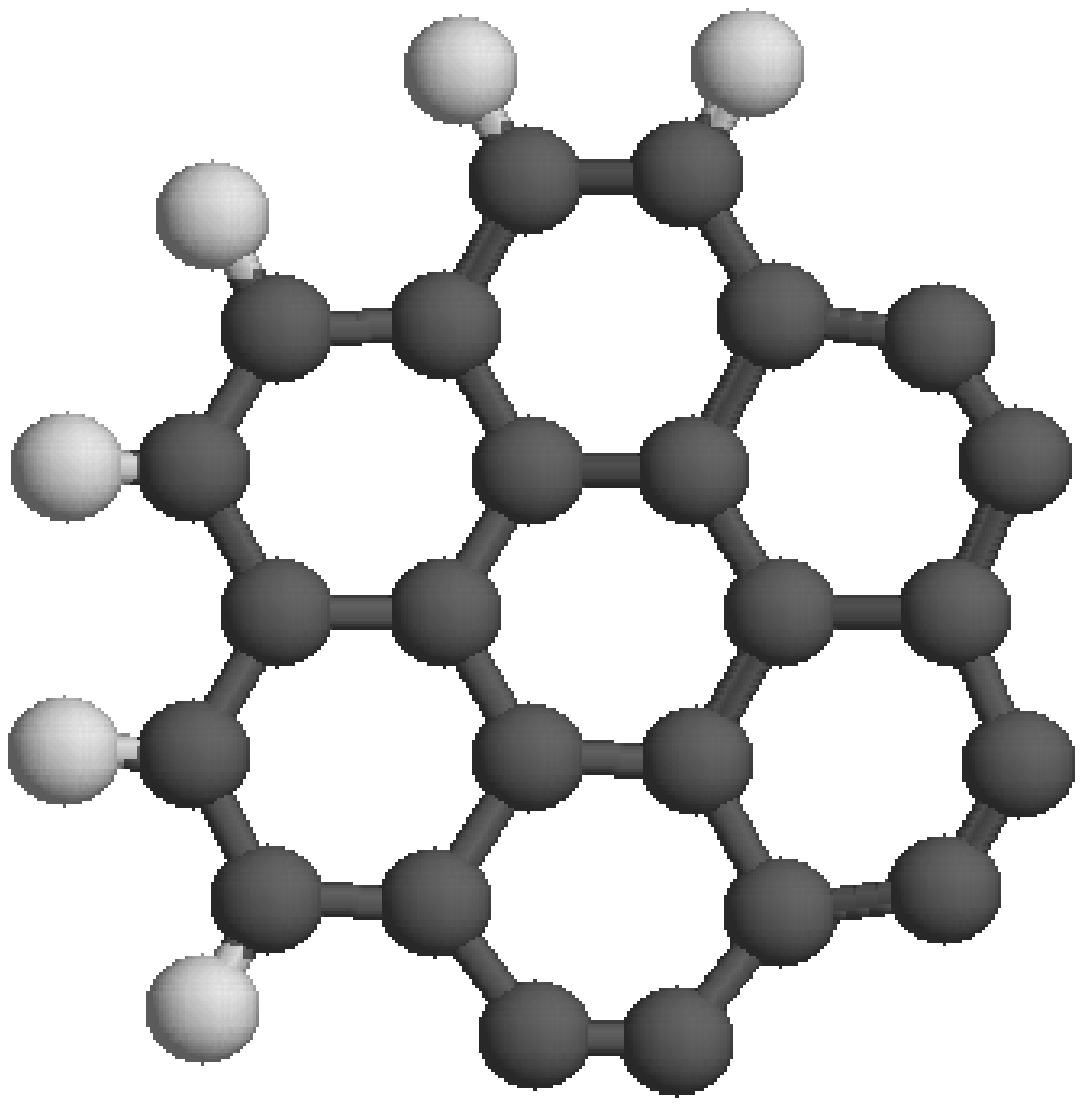}
\includegraphics[trim=3cm 3cm 3cm 3cm, clip,width=2.45cm]{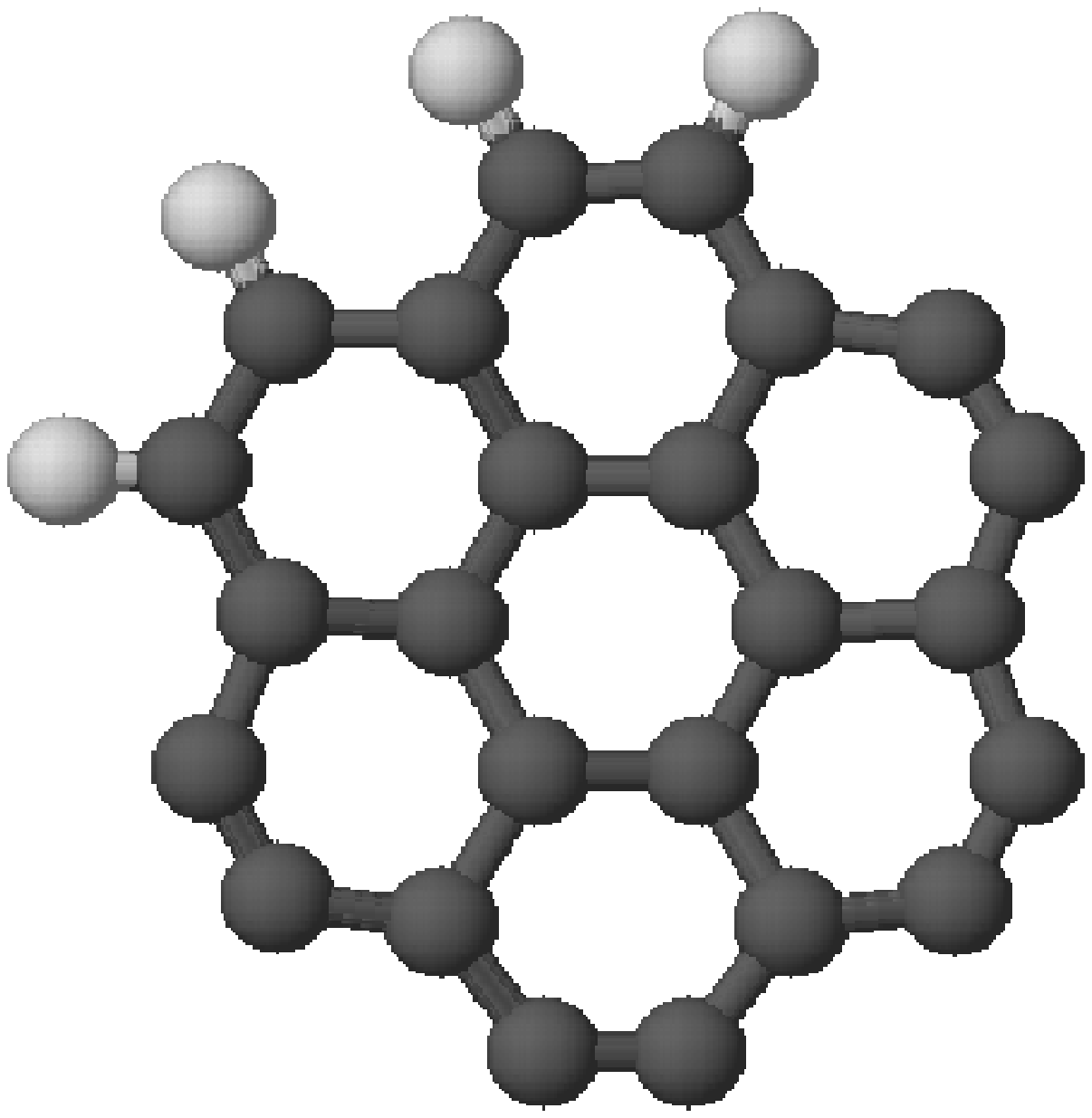}
\includegraphics[trim=3cm 3cm 3cm 3cm, clip,width=2.4cm]{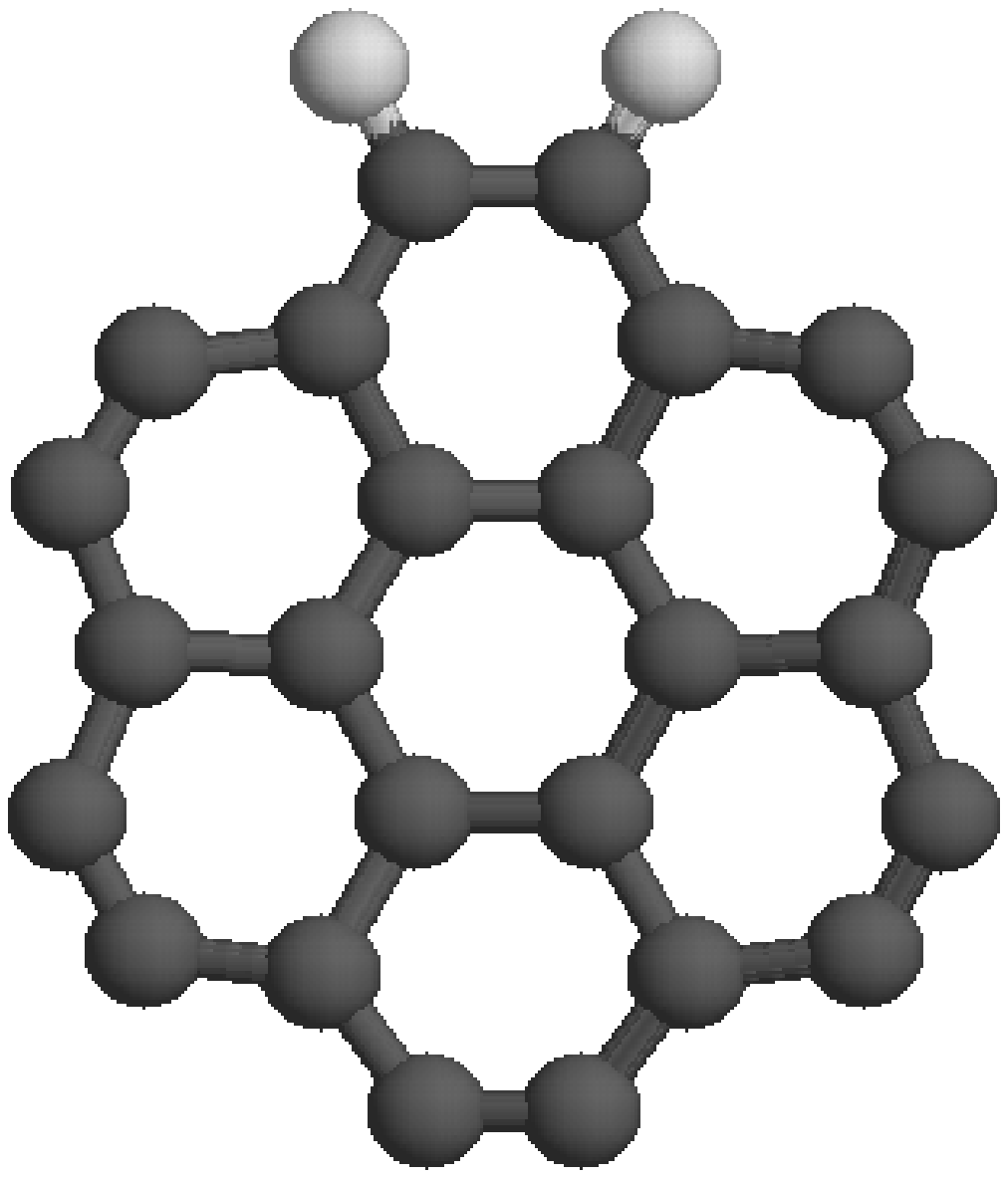}
\includegraphics[trim=3cm 3cm 3cm 3cm, clip,width=2.3cm]{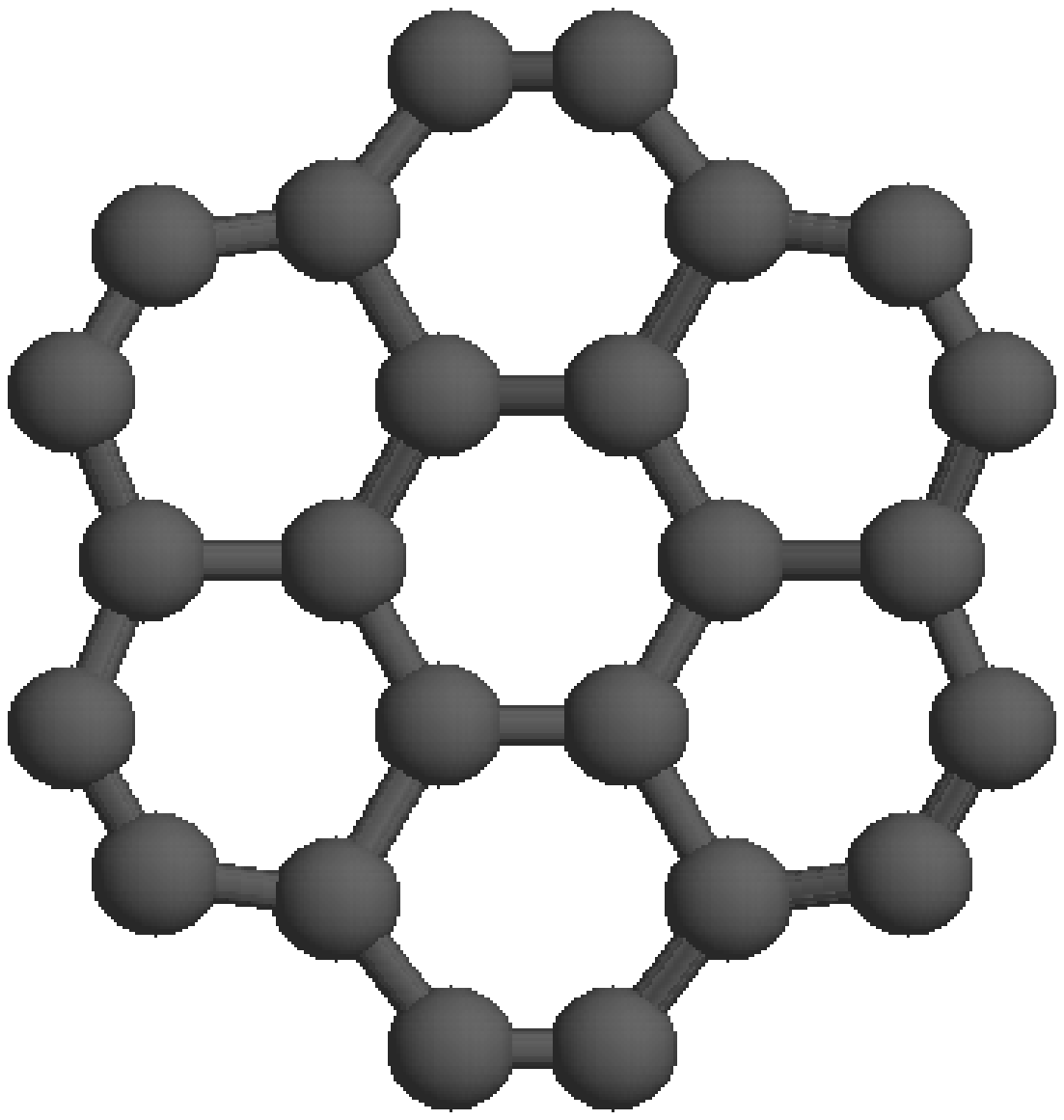}
\includegraphics[trim=3cm 3cm 3cm 3cm, clip,width=2.4cm]{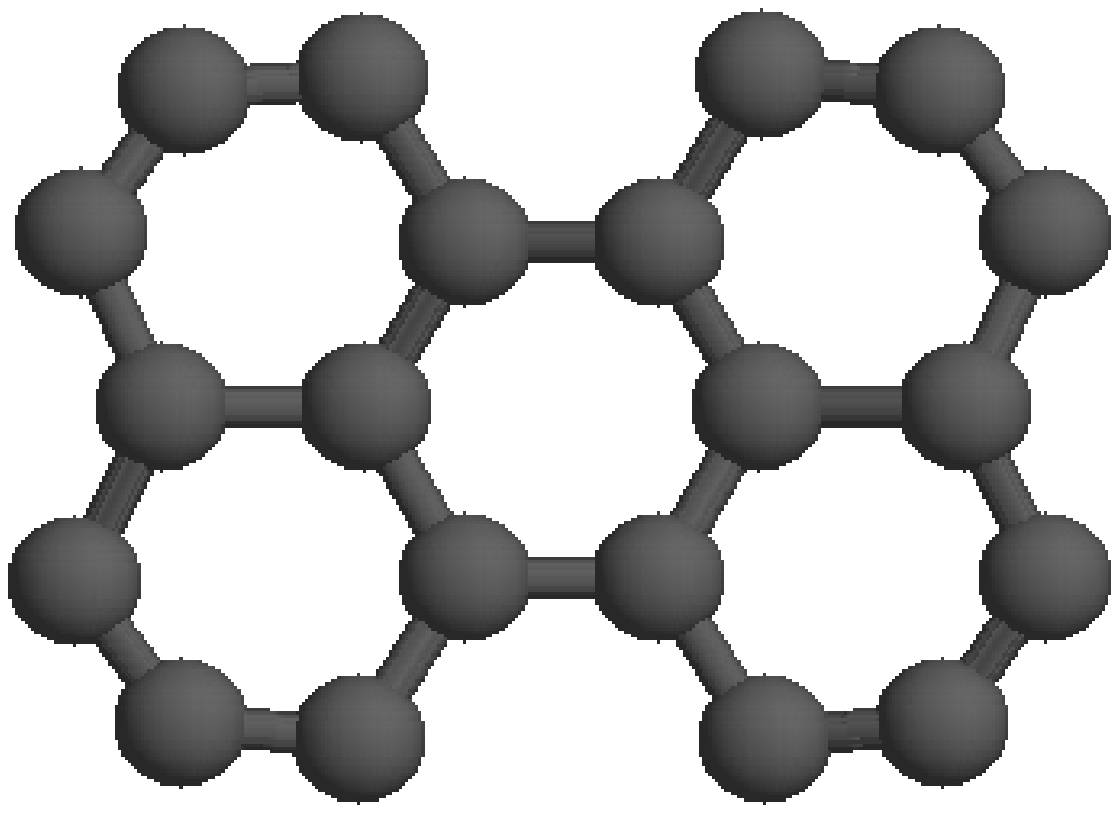}
\includegraphics[trim=3cm 3cm 3cm 3cm, clip,width=2.75cm]{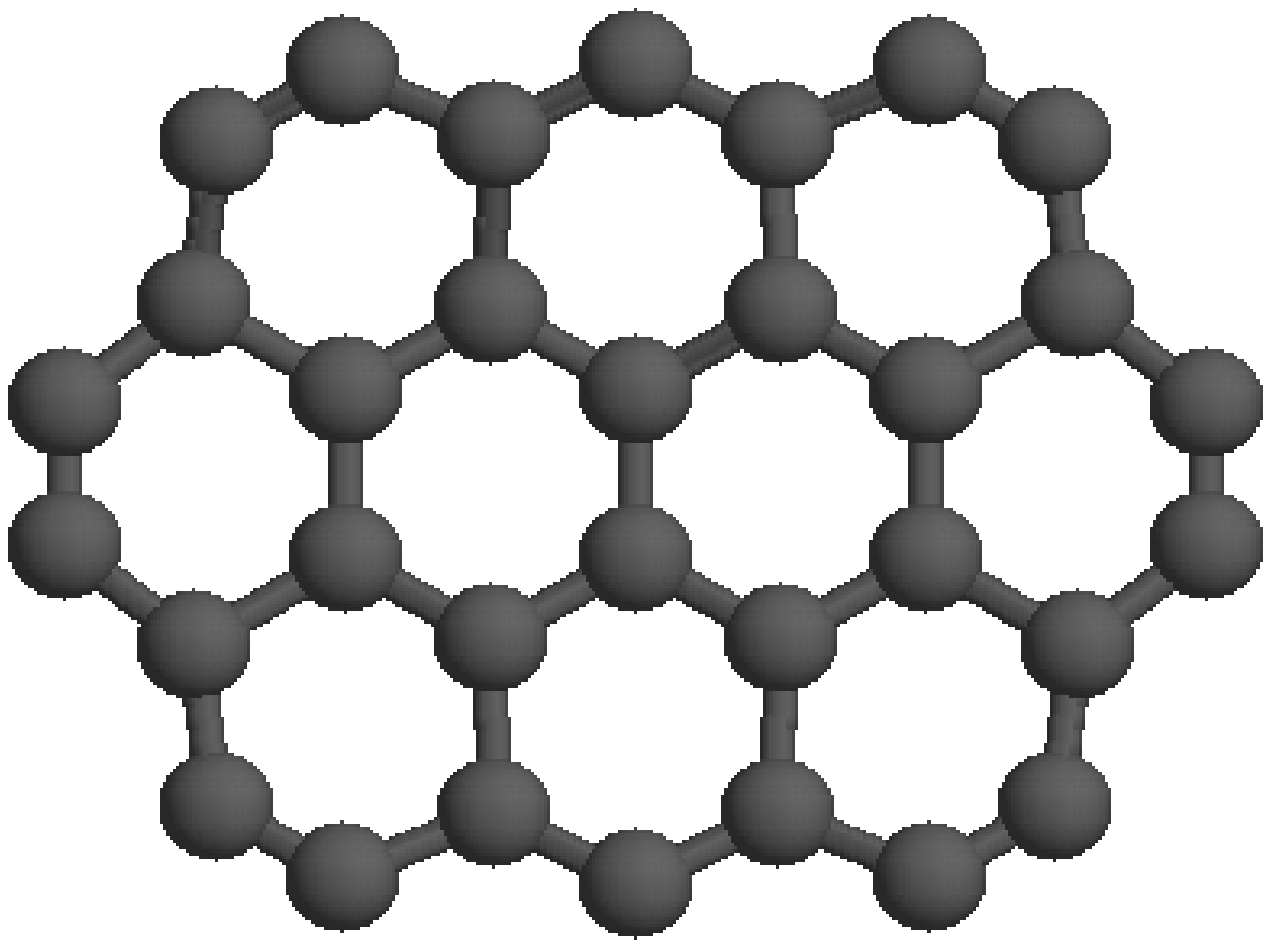}

\caption{Geometries of the dehydrogenated coronene molecules considered
(C$_{24}$H$_{10}$, C$_{24}$H$_{8}$, C$_{24}$H$_{6}$, C$_{24}$H$_{4}$, C$_{24}$H$_{2}$, and  
C$_{24}$), and of the completely dehydrogenated perylene (C$_{20}$)
and ovalene (C$_{32}$).
\label{mols}}
\end{center}
\end{figure}

\section{Methods}\label{methods}
\subsection{Computational details} \label{computation}

As in previous studies \citep{mal07a,mal07b} we used density functional theory 
\citep[DFT,][]{jon89} for the electron ground state, and its 
time\textendash dependent extension \citep[TD\textendash DFT,][]{mar04} for 
electron excited states. Geometry optimisations were first performed using 
the computationally inexpensive \mbox{4\textendash31G} basis set, then 
refined with the \mbox{6\textendash31+G$^\star$} basis 
\citep{fri84}. For this part of the work we used the hybrid B3LYP functional 
\citep{bec93}, as implemented in the Gaussian\textendash based DFT module of 
\textsc{NWChem} \citep{apr05}. 

Starting from the geometry of each fully hydrogenated PAH 
\citep{mal07a}, we considered its dehydrogenated derivatives. 
We considered only neutral and cationic charge\textendash states, and
found in all cases that the computed ground\textendash states for 
neutrals and cations are the singlet and the doublet, respectively. 
All calculations were performed without symmetry constraints. 
To evaluate the effect of dehydrogenation on the electronic structure of the
species considered, we performed exploratory 
Harthree\textendash Fock/\mbox{4\textendash31G} 
calculations. Following Koopman's theorem \citep{ela84}, we interpreted 
the one\textendash electron molecular orbital energies thus obtained as 
vertical ionisation energies. 

Keeping fixed the ground\textendash state optimised geometries obtained above, 
we then computed the absorption cross\textendash section $\sigma(E)$ using 
the real\textendash time TD\textendash DFT implementation of 
\textsc{octopus} \citep{mar03}, in the local\textendash density 
approximation \citep{per81} as described in \citet{mal07a}. We previously 
showed \citep{mal04,mul06b} that the results of this method reproduce the 
overall far\textendash UV behaviour of $\sigma(E)$, as seen in comparison 
with the experimental data available for a few neutral PAHs 
\citep{job92a,job92b}. This includes the broad absorption peak dominated by 
$\sigma\to\sigma^\star$ transitions, which matches well both in position and 
width. The $\pi\to\pi^\star$ transitions from the visible to the UV 
range, while their absolute intensities are well reproduced, show energies 
that are systematically underestimated.

\begin{figure}[h!]
\includegraphics{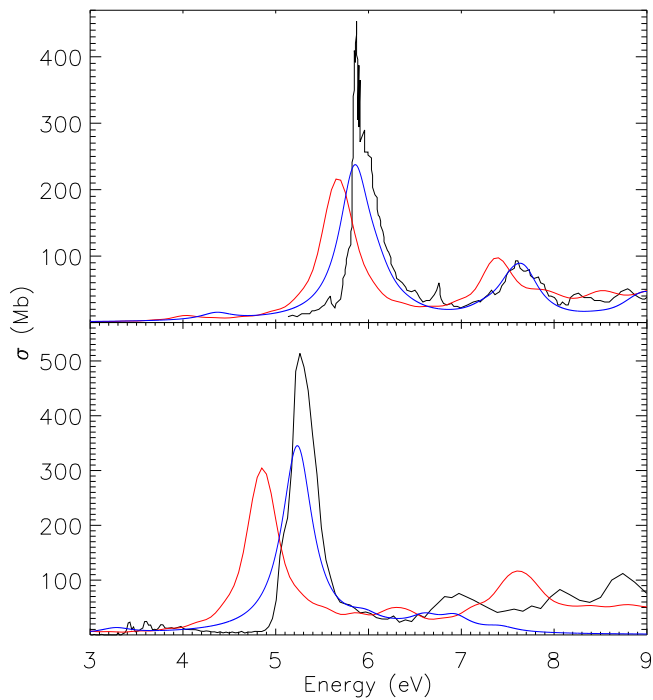} 
\caption{Comparison between the experimental (black line) absorption 
cross\textendash section $\sigma(E)$ in megabarns (1Mb = $10^{-18}$~cm$^2$) of neutral 
naphthalene \citep[top,][]{gin98}, and anthracene 
\citep[bottom,][]{job92a,job92b}, and two different theoretical 
results: the real\textendash time TD\textendash DFT spectrum \citep[red line,][]{mal04}, and the 
frequency space TD\textendash DFT one \citep[blue line,][]{mal07b}.\label{f2}}
\end{figure}

The TD\textendash DFT formalism used in this work yields directly the 
time\textendash dependent linear response of a given molecule after an 
impulsive perturbation, producing the whole spectrum in a single step 
\citep{mar03}. This differs from the most widely used 
frequency\textendash space TD\textendash DFT implementation, where the poles 
of the linear response function correspond to vertical excitation energies,
and the pole strengths to the corresponding oscillator strengths \citep{cas95}.
We compare in Fig.~\ref{f2} the experimental absorption spectra of neutral 
naphthalene \citep{gin98}, and anthracene \citep{job92a,job92b} with both the
real\textendash time TD\textendash DFT \citep{mal04} and the frequency\textendash space TD\textendash DFT \citep{mal07b} 
spectra. The latter results agree better with the laboratory data. However, 
computational costs scale steeply with the number of transitions and electronic
excitations of molecules as large as those considered here are thus 
limited to the visible/near\textendash UV part of the spectrum. 

\subsection{Choice of molecules}\label{sample}

\begin{figure}[h!]
\includegraphics{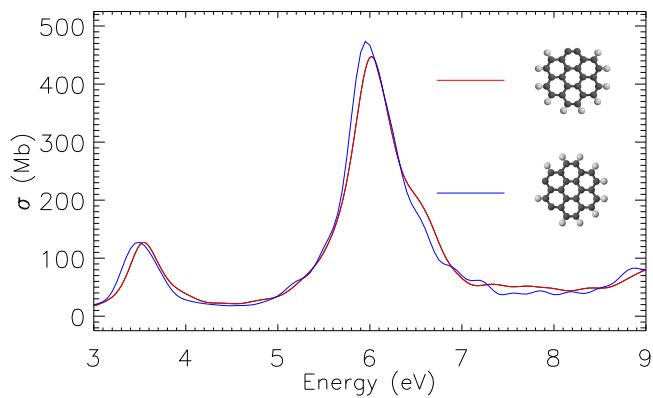} 
\caption{Comparison between the computed absorption cross\textendash section 
of the two inequivalent isomers of C$_{24}$H$_{10}$$^+$ sketched on the right.
\label{f3}
}

\end{figure}

We studied the dehydrogenated derivatives of coronene 
C$_{24}$H$_{n}$ (n=10,8,6,4,2,0) in their neutral and cationic charge states.
We included only species with an even number of hydrogen atoms, based on 
experimental mass spectra \citep{eke97} showing that dehydrogenation 
corresponding to loss of an even number of hydrogen atoms is dominant. 
In addition, experimental results on the dehydrogenation of the coronene 
cation \citep{job08} show that, for C$_{24}$H$_{2n}^+$ molecules,
only species containing adjacent paired hydrogens have to be considered.
In particular, \citet{job08} found that the main hydrogen loss channel is 
the detachment of single atoms; however, the photodissociation rate
for ``lone'' hydrogen atoms was observed to be much faster (about one order of
magnitude) than for paired hydrogen atoms. This implies that dehydrogenated 
coronene derivatives containing adjacent paired hydrogens are more likely to 
be found in interstellar conditions. Moreover, it appears that the placement 
of the hydrogens at the periphery of the carbon skeleton has a minor effect on 
the main $\pi \to \pi^\star$ transitions. Such transitions depend mainly on the number 
of electrons populating the resonant $\pi$ and $\pi^\star$ molecular orbitals, which 
depend only on the number of peripheral hydrogen atoms and not on their 
specific position. As an example, Fig.~\ref{f3} shows the comparison between 
the computed absorption cross\textendash sections of two isomers of C$_{24}$H$_{10}$$^+$,
whose geometries are sketched in the same figure. As expected, with the 
exception of small details, the spectra of the two isomers are found to be 
almost indistinguishable in the energy range of astrophysical interest. Our 
study can therefore be restricted, without loss of generality for the scope of 
the present paper, to only the isomers of C$_{24}$H$_{n}$ (n=10,8,6,4,2,0) shown 
in Fig.~\ref{mols}. We additionally examined the completely dehydrogenated 
derivatives of perylene (C$_{20}$H$_{10}$) and ovalene (C$_{32}$H$_{14}$) cations, 
i.~e. C$_{20}^+$ and C$_{32}^+$, representative of non\textendash symmetric, dehydrogenated, 
compact, medium\textendash sized PAHs.

\section{Results and discussion} \label{results}

\subsection{Dehydrogenation and electronic excitation properties} 

\begin{figure}
\includegraphics{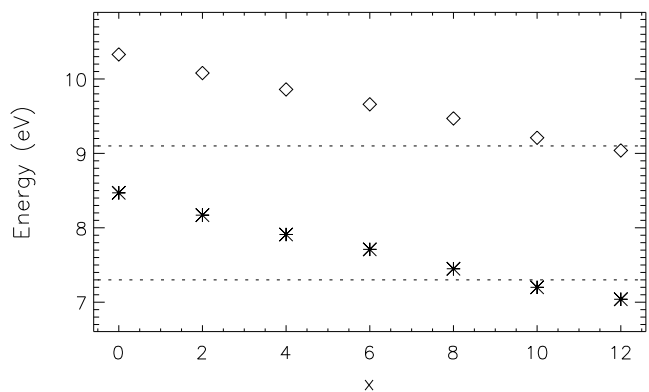}
\caption{First (asterisks) and second (diamonds) vertical ionisation energies 
of dehydrogenated coronene molecules C$_{24}$H$_{x}$, computed at the 
Harthree\textendash Fock/\mbox{4\textendash31G} level in the framework of 
Koopman's theorem, as a function of the degree of hydrogenation. 
The horizontal dotted lines correspond to the low\textendash{} and 
high\textendash frequency cutoffs extracted from the interstellar 
extinction curve in the model by \citet{dul06a}.
\label{f4}
}
\end{figure}

According to Koopman's theorem the molecular orbital energies obtained at 
the Harthree\textendash Fock / \mbox{4\textendash31G} level are interpreted 
as vertical ionisation energies. We report in Fig.~\ref{f4} the energies 
corresponding to removal of an electron from both the highest occupied 
molecular orbital (IP1) and the molecular orbital just below it (IP2), 
as a function of the degree of hydrogenation ($x=0,2,4,6,8,10,12$). In 
the model put forward by \citet{dul06a} the low\textendash{} and 
high\textendash frequency cutoffs extracted from the interstellar
extinction curve, at about 7.3 and 9.1~eV respectively, are identified 
with the IP1 and IP2 of the carrier of the UV bump. These values
were found to be in close agreement with those measured for neutral
coronene, 7.29 and 9.13~eV, respectively \citep{cla81}. Based on the 
assumption that $\pi$\textendash electron energies are little affected by 
dehydrogenation, this supported the proposal that a $\pi\to\pi^*$ plasmon 
resonance in neutral and singly\textendash ionised dehydrogenated coronene 
molecules could generate the UV bump. However, we found an approximately 
constant change in ionization energy per hydrogen atom of about 0.1~eV 
(cf. Fig.~\ref{f4}), suggesting that $\pi$\textendash electron 
energies are indeed strongly affected by dehydrogenation. This is confirmed for
the adiabatic first ionization energies computed via total energy differences 
at the B3LYP/\mbox{6\textendash31+G$^\star$} level: 7.02, 7.23, 7.38, 7.61, 
7.79, 8.00, and 8.25~eV, when going from C$_{24}$H$_{12}$ to C$_{24}$. 

\begin{figure}
\includegraphics{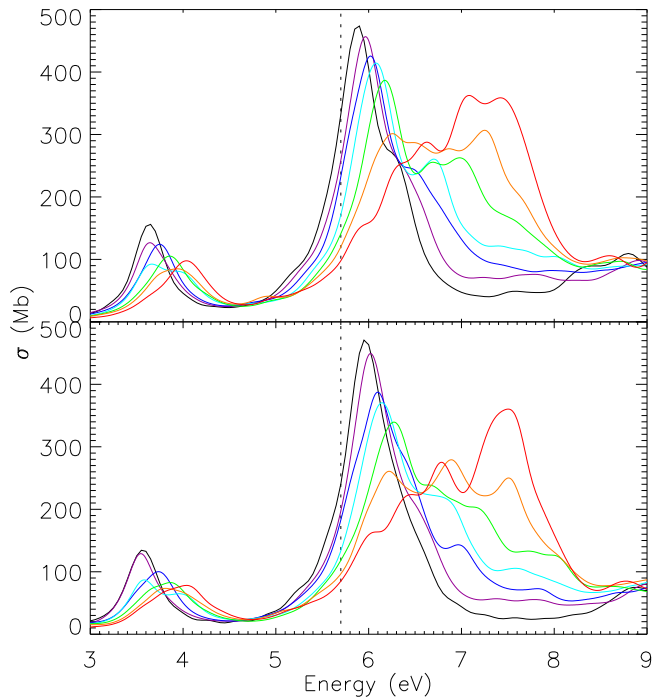}
\caption{Comparison between the computed absorption 
cross\textendash sections of coronene C$_{24}$H$_{12}$ (black), C$_{24}$H$_{10}$ (violet), 
C$_{24}$H$_{8}$ (blue), C$_{24}$H$_{6}$ (cyan),  C$_{24}$H$_{4}$ (green), 
C$_{24}$H$_{2}$ (orange), and C$_{24}$ (red). Neutral molecules and their 
corresponding cations are displayed in the top and bottom panel, respectively. 
The dotted vertical line represents the position of the observed 
UV bump at $\sim$5.7~eV.\label{f5}}
\end{figure}

\begin{figure}
\includegraphics{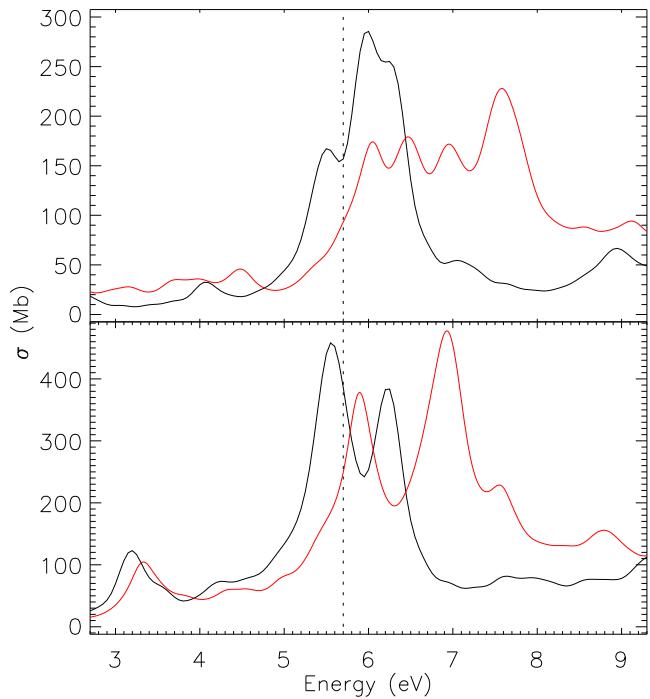}
\caption{Comparison between the computed absorption cross\textendash sections of
fully hydrogenated (black), and fully dehydrogenated (red) perylene 
cation (C$_{20}$H$_{10}^+$) (top panel) and ovalene cation (C$_{32}$H$_{14}^+$) 
(bottom panel). \label{f6}}
\end{figure}

The comparison between the computed absorption cross\textendash sections of
the different dehydrogenated coronene molecules considered, in their neutral 
and cationic charge\textendash states, is presented in Fig.~\ref{f5}. The corresponding
comparisons between C$_{20}$H$_{10}^+$ and C$_{32}$H$_{14}^+$ and their 
fully dehydrogenated counterparts are shown in Fig.~\ref{f6}. For both 
neutral and singly\textendash ionised coronene 
derivatives, progressive dehydrogenation translates into a correspondingly 
progressive blue shift of the main electronic transitions (Fig.~\ref{f5}). The 
$\pi\to \pi^\star$ collective resonance, in particular, becomes broader as 
well as bluer with dehydrogenation. The same holds true for  
the other PAHs reported in this work,
namely perylene and ovalene (Fig.~\ref{f6}), suggesting that this is a 
general feature of the whole class of molecules.

\subsection{Astrophysical implications}

\begin{figure}
\includegraphics{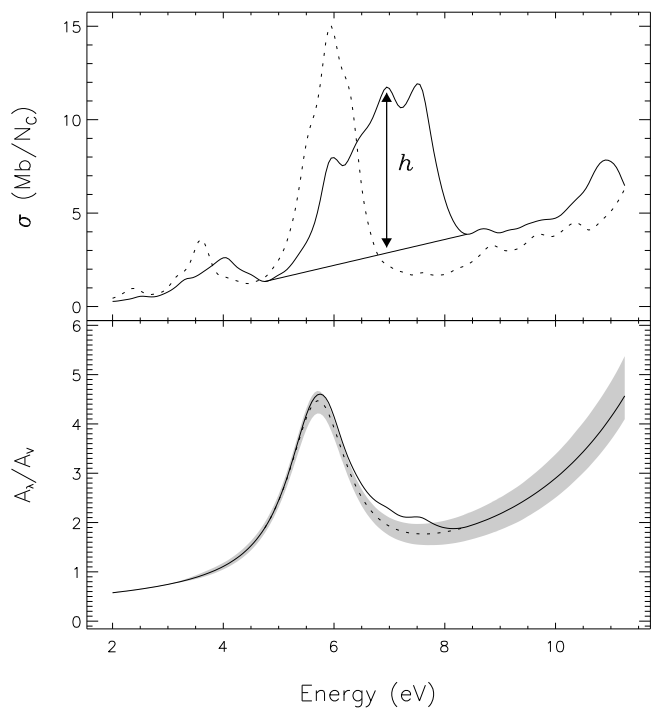}
\caption{Top panel: 
Weighted sum of the computed absorption cross\textendash sections of 
C$_{20}^+$, C$_{24}$, C$_{24}^+$, and C$_{32}^+$ (continuous line), 
and C$_{20}$H$_{10}^+$, C$_{24}$H$_{12}$, C$_{24}$H$_{12}^+$, and 
C$_{32}$H$_{14}^+$ (dotted line), using as weight the inverse of the 
number of C atoms in each molecule. 
Bottom panel: comparison between the average extinction curve 
\citep[dotted line,][]{fit07}, with the 1$\sigma$ region in 
gray shade, and the average extinction curve + the amount of 
dehydrogenated PAHs which would make the $\pi\to\pi^\star$ feature detectable 
at 2$\sigma$ (continuous line). 
\label{weighted}}
\end{figure}

For a quantitative comparison with extinction, we estimate the expected 
spectral contrast of the  $\pi\to \pi^\star$ collective 
resonance in a mixture of all the fully dehydrogenated species 
considered so far. The top panel of Fig.~\ref{weighted}
shows the weighted sum of their absorption spectra, compared with the
same weighted sum for their fully hydrogenated parent molecules.
The $\pi\to \pi^\star$ feature appears to peak at $\sim$6.95~eV
($\sim$178~nm), with an integrated intensity above the underlying
continuum of about 1.3 times larger than the strong $\pi\to \pi^\star$ feature
at 5.95~eV associated with the hydrogenated counterparts.

\citet{ccp08} showed that mixtures of a large number of 
fully hydrogenated PAHs can reproduce the UV bump and 
non\textendash linear far\textendash UV rise of the extinction curve. 
In particular, the UV bump is very accurately matched in position 
($\sim$5.7~eV), intensity and width. Such a fit can only be obtained using 
mixtures including a much larger sample of PAHs than those included here. 
However, the position of the $\pi\to \pi^\star$ resonance, and its intensity
per carbon atom, are relatively insensitive to the individual species 
considered. 
In all cases, its peak is well within the observed width of the UV bump. 
This is the reason why such transitions, in a mixture of 
hydrogenated PAHs, become blended, yielding a smooth feature which can be very 
similar to the observed UV bump profile. 

In contrast, all completely dehydrogenated species considered here have the 
peak of their $\pi\to \pi^\star$ resonance about $\sim$1~eV bluewards of the bump, in the 
gap between it and the onset of the non\textendash linear far\textendash UV rise. The position and 
intensity per carbon atom of the $\pi\to \pi^\star$ resonances of dehydrogenated PAHs, 
as a family, can be expected to be as insensitive to individual species as for 
their fully hydrogenated counterparts. A larger mixture, if this is the case, 
would yield a smoother feature than the one we calculated with the present 
sample, but generally with the same intensity and position.

Since no structure at $\simeq7$~eV is detected in the average galactic
interstellar extinction curve reported by \citet{fit07}, fully dehydrogenated 
PAHs must produce a peak smaller than the error of the interstellar
extinction curve as illustrated in Fig.~\ref{weighted}.
Using a $2\sigma$ upper limit, we can derive that the integrated intensity 
of the $\pi\to \pi^\star$ feature of dehydrogenated PAHs is less than $1/6$ 
of the bump area. Assuming that the bump is mainly due to hydrogenated PAHs 
\citep{ccp08} and using the factor given above for the relative integrated 
$\sigma(E)$, we derive that less than 1/8 of the total abundance of carbon 
in PAHs is contained in strongly dehydrogenated PAHs.
Assuming $\mathrm{N}_\mathrm{H} \simeq 5.9 \times 10^{21} 
\mathrm{E}_\mathrm{B-V}$ \citep{wit02}, and applying the procedure 
of \citet{ccp08} to the average extinction curve 
\citep[$\mathrm{R}_\mathrm{V} \simeq 3.01$ from][]{fit07} yields a value of 
$\sim$150~ppM for the abundance of C locked in PAHs (including both the 
free\textendash flying ones and those clustered in very small grains);
the upper limit of the contribution of dehydrogenated PAHs
to the extinction curve can then be translated into an upper limit of 
$\sim$18~ppM for the carbon atoms contained in them. This numeric result is 
obviously dependent on the choice to compare against the average 
interstellar extinction curve, and would be different for a different 
choice. We showed in \citet{ccp08} that the fraction 
of carbon contained in free\textendash flying and clustered PAHs 
is quite variable, from 85 to 187 ppm in a sample of lines of sight.

Our upper limit can be compared with those found by \citet{cla03} regarding 
the column density of specific small, neutral, fully hydrogenated PAHs in 
specific lines of sight, based on laboratory spectra and Hubble Space Telescope
observations. Their results are more stringent thanks to the stronger spectral
contrast of the absorption bands in their experimental spectra, coupled with
the matching high resolving power of the STIS instrument. For the present 
comparison, however, such higher resolving power would not help, given the
broader nature of the spectral signature being sought.

\section{Conclusions}
\label{conclusions}

PAHs are expected to exist in a wide variety of interstellar environments,
in a complex statistical equilibrium of different charge and hydrogenation
states \citep[see e.g.,][]{tie05}. Modelling studies suggest that 
intermediate\textendash size PAHs in the range of 20\textendash30 carbon atoms are stripped of 
most of their hydrogen atoms but still survive under the conditions of the 
diffuse interstellar medium \citep{lep03}.

We undertook a systematic theoretical study of the effects of dehydrogenation
on the electron properties of neutral and ionised PAHs using 
state\textendash of\textendash the\textendash art quantum\textendash chemical 
techniques in the framework of the real\textendash time TD\textendash DFT. 
For all of the species considered so far, we found that, regardless of 
charge\textendash state and molecular size, the $\pi\to \pi^\star$ collective 
resonance broadens and shifts to higher energies with increasing 
dehydrogenation. 

When the molecule is half dehydrogenated or more, the band is shifted enough 
to fall between the bump and the far\textendash UV rise of the extinction 
curve. The associated species cannot be the carriers of the bump, 
contrary to the proposal of \citet{dul06a}.

While it is overreaching to draw definitive conclusions based on the few cases 
considered, our upper limits are due to systematic effects, and are thus 
unlikely to depend strongly on the specific sample of molecules considered. 
Nonetheless, a larger study, which is underway \citep{mal08}, will be needed 
to assess to what extent the effect of dehydrogenation that we observed in UV 
spectra of PAHs is systematic for the whole class.

\begin{acknowledgements}
G.~Malloci acknowledges financial support by Regione Autonoma della
Sardegna. G.~M., G.~M., and C.~C.\textendash P. acknowledge financial support 
by MIUR under project CyberSar, call 1575/2004 of PON 2000\textendash2006. We 
acknowledge the authors of \textsc{octopus} and the authors of 
\textsc{nwchem}, A Computational Chemistry Package for Parallel Computers, 
Version~4.7 (2005), PNNL, Richland, Washington, USA. 
Parts of these simulations were carried out at CINECA (Bologna).

\end{acknowledgements}

\end{document}